\documentclass{article}
\usepackage[final]{colm2026_conference}

\usepackage{microtype}
\usepackage{hyperref}
\usepackage{url}
\usepackage{booktabs}
\usepackage{graphicx}
\usepackage{xcolor}
\usepackage{tikz}
\usetikzlibrary{positioning, arrows.meta, calc, fit}
\usepackage{lineno}

\usepackage{twemojis}
\newcommand{\AffGT}{\twemoji{bee}}          
\newcommand{\AffGTSafety}{\twemoji{shield}} 
\newcommand{\AffMATS}{\twemoji{fire}}       

\definecolor{darkblue}{rgb}{0, 0, 0.5}
\hypersetup{colorlinks=true, citecolor=darkblue, linkcolor=darkblue, urlcolor=darkblue}

\newcommand{\reffig}[1]{Figure~\ref{fig:#1}}
\newcommand{\reftab}[1]{Table~\ref{tab:#1}}
\newcommand{\refsec}[1]{Section~\ref{sec:#1}}
\newcommand{\refapp}[1]{Appendix~\ref{app:#1}}
\newcommand{\wopr}{\textsc{WOPR}}

\newcommand{\blfootnote}[1]{%
  \begingroup
    \renewcommand{\thefootnote}{}%
    \footnote{#1}%
    \addtocounter{footnote}{-1}%
  \endgroup
}

\definecolor{MLBlue}{HTML}{2F6FA3}
\definecolor{MLGreen}{HTML}{2E7D4F}
\definecolor{MLAmber}{HTML}{B36B00}
\definecolor{MLPurple}{HTML}{6A4C93}
\colorlet{enginecolor}{MLBlue}
\colorlet{agentcolor}{MLAmber}
\colorlet{phasecolor}{MLGreen}

\title{No One Wins in Nuclear War\\
\large A Social Simulation of Military Decision-making}

\author{%
  Glenn Matlin$^{\AffGT\,\AffGTSafety\,\AffMATS}$ \\
  Isaac Song$^{\AffGT\,*}$ \quad Anthony Wen-Ming Zang$^{\AffGT\,\AffGTSafety\,*}$ \\
  Mark Riedl$^{\AffGT}$ \\[4pt]
  \AffGT\,College of Computing, Georgia Institute of Technology \\
  \AffGTSafety\,Georgia Tech AI Safety Initiative \quad
  \AffMATS\,MATS Program
}

\begin{document}
\ifcolmsubmission\linenumbers\fi
\maketitle
\lhead{Published at the Social Sim'26 Workshop at COLM 2026}
\blfootnote{\textsuperscript{*}Equal contribution.}

\begin{abstract}
\wopr{} is a social-simulation environment for studying how organizations make
high-stakes decisions, built on a deterministic, replay-validated rules engine
and using wargames as the vehicle. We instantiate it first with the published
card game \emph{Nuclear War}, traced against its published rules. We start with
military decision-making because of its safety implications and because it
needs further study, but the design is not specific to it: the decision-point
contract that exposes the engine to agents is reusable across verifiable rule
systems. Existing social-simulation work emphasizes persona fidelity and
synthetic opinion, but lacks a verifiable rules engine with replay-checkable
mechanics and private-channel negotiation. \wopr{} supplies that engine, and
its contract makes every strategic choice an explicit agent decision. The
method is agnostic to social-simulation frameworks; we adopt Concordia as the
default harness for driving the game. On the same engine, \wopr{} layers a
four-rung press ladder from silence to private single-recipient channels with
structured commitments, and instantiates each faction as a collective
command-and-control system rather than a single agent.
We make all code, example configurations, and replay data publicly available at
\url{https://github.com/eilab-gt/wopr}.
\end{abstract}

\section{Introduction}\label{sec:intro}
Recent social-simulation work targets persona fidelity, belief-behavior alignment
and synthetic opinion
formation \citep{park2023generative,simbench2025,socialmaze2025}, but rarely
places consequences as the object of study.

We focus on \emph{war decisions}: strategic, adversarial choices whose
consequences must be checkable. To study them rigorously, simulations require
a game engine that owns the rules, supplies the legal actions, and records
every state transition. Any claimed outcome can be replayed and audited.
We aim to supply tooling for organizational decision-making.

No current social-simulation environment provides such an engine. The same
gap appears in matrix and open-ended wargaming with language models, where a
model or human adjudicates mechanics instead of a deterministic engine and no
shared decision contract binds external agents
\citep{lamparth2024wargame,openended_wargames_2024}.

We present \wopr{}, a social-simulation and causal-analysis environment for studying organization-level, high-stakes decision-making. The environment is
built on a deterministic, replay-validated rules engine and is intended for
wargames more generally; we instantiate it first with \emph{Nuclear War}, a published card game from Flying Buffalo, with implemented mechanics traced to source in a conformance matrix (\refapp{conformance}). Any player that satisfies the
contract can act as an agent; the engine is the sole source of legal actions
and records every transition, so any game can be replayed and audited. Our
contributions are:
\begin{itemize}
  \item A simulation environment built on a replay-validated rules engine,
  instantiated on a real published game and verified against a 240-game sweep
  that confirms seeded outcomes are preserved across changes to the engine
  (\refsec{testbed}).
  \item A decision-point contract that turns every strategic choice into an
  explicit agent decision at a deterministic, rule-defined point in the turn
  (\refapp{contract}).
  \item A four-rung press ladder that exposes communication capacity as a
  variable on an unchanged engine, from silence through public multi-turn
  exchange to full press with private channels and structured commitments
  (\refsec{press}).
  \item Four structurally distinct faction command-and-control archetypes,
  each motivated by a real-world nuclear-use command system, all of which
  converge on one release action per decision on the same contract
  (\refsec{faction}).
\end{itemize}

This is an early-stage paper: we present the environment, not a controlled
study.

\section{Background and Related Work}\label{sec:bg}
A first line of work pursues believable agent personas and the social dynamics
they produce. \citet{park2023generative} introduced generative agents that
remember, reflect, and plan in a small-town sandbox. That work established the
memory-and-retrieval architecture that much of the field now builds on. The
Concordia library \citep{vezhnevets2023concordia} generalizes that idea into a
harness for generative agent-based models in physical, social, or digital
spaces.

A second line turns those agents into measurable benchmarks of social
behavior. SimBench \citep{simbench2025} aggregates diverse datasets to evaluate
how faithfully language models simulate group-level human behavior.
SocialMaze \citep{socialmaze2025} probes social reasoning under information
uncertainty and deception. NegotiationGym \citep{negotiationgym2025} provides a
configurable multi-agent simulation for self-optimizing agents in negotiation.
A related thread studies governance and collective choice, including
cooperation over shared resources \citep{piatti2024govsim} and elections in
AI societies \citep{deshpande2025govsimelect}.

What none of these lines supply, for our purposes, is a verifiable and
replay-checkable engine for the rules that govern the simulated choices. A
model or an adjudicator produces the consequences instead of auditable
transitions.

A separate line of work applies language models to wargaming, crisis
simulation, and negotiation-heavy strategy games
\citep{rivera2024escalation,hua2023waragent,bakhtin2022diplomacy}.
\citet{lamparth2024wargame} compare expert humans and language
models in a crisis-escalation wargame and report systematic behavioral
differences, including a tendency of models to escalate. Open-ended wargames
with language models \citep{openended_wargames_2024} automate qualitative
wargaming with a multi-agent system.

These lines share the same gap: the mechanics are not encoded in an engine
whose transitions an external observer can replay. Without that, researchers
cannot replay the same game to check whether a claimed outcome follows from
the stated rules, nor compare agents across communication conditions while
holding the rules fixed.

\wopr{} belongs to the benchmark and environment cluster exemplified by
SimBench, SocialMaze, NegotiationGym, and GOVSIM-ELECT. It differs along two
axes. First, rule conformance: the rules are a deterministic implementation
of a real published game, and any game is replay-checkable against them
(\refapp{conformance}). Second, reproducibility: the decision-point contract
fixes what an agent must satisfy, and the press ladder fixes the
communication axis. Researchers can therefore compare behavior across agents
and communication conditions on an unchanged rules engine.

\section{The WOPR Testbed}\label{sec:testbed}
\begin{figure*}[t]
\centering
\includegraphics[width=0.98\linewidth]{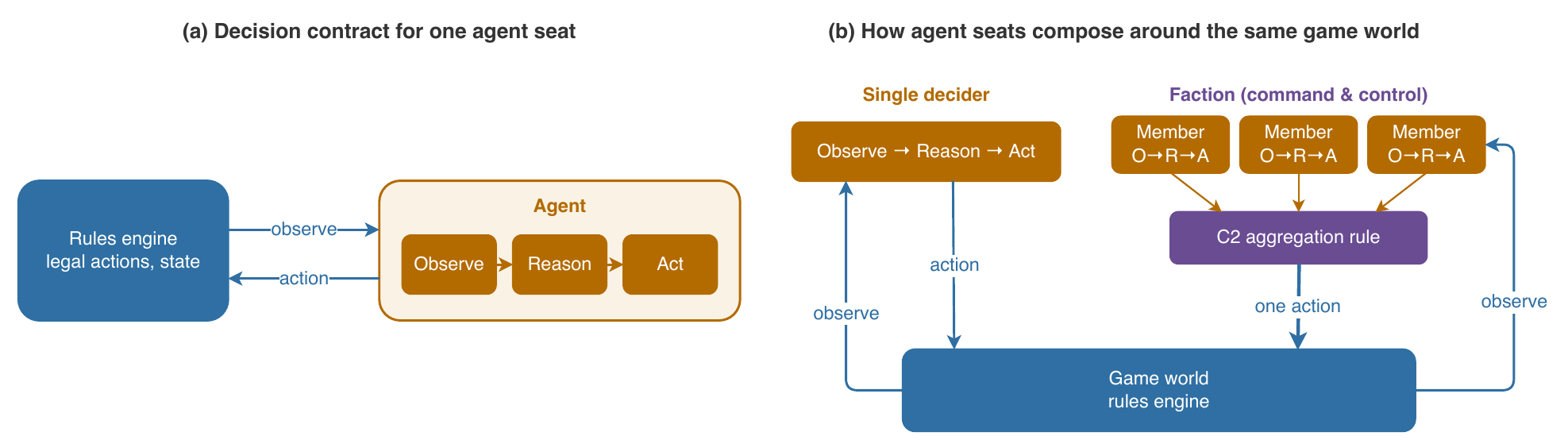}
\caption{(a)~The decision-point contract for a single agent seat: the engine remains the
sole and authoritative source of legal actions, with the agent observing, reasoning, and then acting by selecting among them. (b)~How agent seats compose around the same game world: a single decider runs one observe-reason-act loop, while a faction may run one for each member, following it with a command-and-control rule to collapse the candidates into one
concrete action. Both seats return one \texttt{LegalAction} per decision, so
they are interchangeable in the same game.}
\label{fig:agent-organization}
\end{figure*}

The engine exposes a small, fixed contract for agent decisions: it reports the
current choice, applies the chosen action, and advances mandatory steps
until the next genuine choice (\refapp{contract}). Individual personas and
multi-agent decision systems are both valid players; the latter need not be
visible to the engine. \reffig{agent-organization} shows how one or many agents
compose against the same world; \refsec{faction} instantiates that design as
per-faction command systems. The engine alone supplies legal actions, and
observation preserves hidden information.

\begin{figure}[t]
\centering
\includegraphics[width=0.92\linewidth,height=95pt,keepaspectratio]{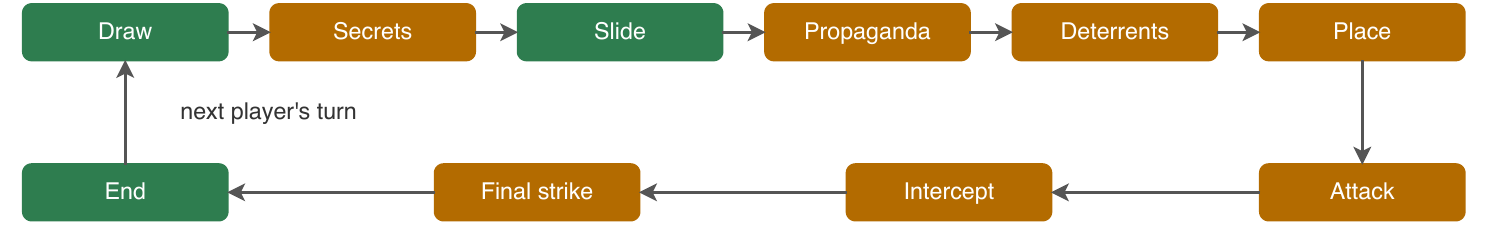}
\caption{The turn-phase cursor shown as a deterministic cycle. 
Mandatory steps (green) advance automatically; genuine choices (amber) pause for an agent decision. On End, the cursor passes to the next player.}
\label{fig:phase-cursor}
\end{figure}

A turn is a sequence of rule-defined phases in which mandatory steps advance
automatically and only genuine choices pause as agent decisions. This turns
the rulebook's turn structure into a deterministic state machine, so the
order and kind of agent choices are fixed by the rules (\reffig{phase-cursor}).
The engine surfaces several strategic decision types (for example, card
placement, launch targeting, and defender interception), each an agent
decision on the same cursor. The phase inventory and decision catalog are in
\refapp{contract}.

Because the engine is deterministic, researchers can replay any game and check
it against the rules. We use replay validation as a fidelity method, anchored
on one principle: purely structural changes to the engine must leave seeded
outcomes unchanged, and any change to outcomes must be explicitly asserted.
This is checked across a 240-game sweep (3 and 4 players, heuristic and random
agents, seeds 1--60). \refapp{replay} details the method.
The paper-facing demo captures are collected in
\refapp{demo-visualizations}: \reffig{demo-replay-table-event} shows the replay
workbench, and \reffig{demo-live-play} shows the same interface connected to a
local live rules API.

\section{The Press Ladder}\label{sec:press}
The \emph{press ladder} adds a communication axis to the same verified-rules
engine: four rungs that increase what players may say while leaving the rules
unchanged (\reffig{press-ladder}). Speech is trace-only, never mutates game state, and leaves the
replay JSON identical across rungs. The same game can therefore run at
different rungs (communication constraints), so any effect of what players may say is measured against a
fixed engine.

In \emph{no-press}, agents observe only their own game state. In
\emph{press-light}, each living agent may make one public statement per round.
In \emph{multi-turn public}, each round runs a fixed $N$ communication passes
in which every living agent speaks once per pass and sees the accumulating
transcript.

The fourth rung, \emph{full press}, adds private single-recipient channels and
structured commitments to the public exchange. A commitment is recorded as
data and linked to later decisions, with no automatic violation detection, so
it is an honor-system signal analyzed after the game. Per-pass option
mechanics and the privacy rule are in \refapp{pressmech}.

\section{Collective Decision-Making: Factions as Command Systems}\label{sec:faction}
The contract of \refsec{testbed} treats any agent uniformly, and a player
need not be a single persona. Each faction is in practice a small
\emph{command-and-control} (C2) system rather than a single agent. The C2
system maps its members' inputs to one release action. The decision-point
contract is unchanged; a faction is a composite agent that runs an internal
deliberation and returns one action.
Externally it is indistinguishable from a single agent, so the same engine and
the same replays apply regardless of which C2 structure is active.

We define four archetypes along one literature-supported axis: where release
authority aggregates and whether fresh deliberation occurs at release
(\reffig{faction-c2}). A
\textbf{sole-authority} faction concentrates authority in one executive
who consults a staff but may ignore it. A \textbf{council} must concur:
members vote, a threshold binds the body, and the executive cannot
override. A \textbf{distributed} faction pre-delegates release to several
holders, any of whom may act. An \textbf{automated} faction removes
fresh deliberation entirely: a policy armed in advance fires on a
trigger. \reftab{c2origins} (\refapp{factionc2}) motivates each archetype with
a real-world nuclear-use command system as its origin, not a fidelity target;
the configuration each exposes is listed in \reftab{c2params}.

\section{Discussion and Future Work}\label{sec:discussion}
Two study families follow directly. The first varies the press rung as the
independent variable, holding the engine fixed, to measure communication
effects. The second varies C2 archetype and personality parameters across
faction command systems. Both require games auditable after the fact, which
the engine's recorded transitions and per-agent views provide. Beyond those,
the decision-point contract is not specific to \emph{Nuclear War}, so the
environment extends to other verifiable rule systems.

The scope here is deliberately narrow: there is no controlled comparison
across rungs and no claim about how agents behave under the press ladder. Press modes are single-game and are not
yet evaluated at scale; multi-game batch analysis is left to future work.
Commitments are recorded but not enforced. Of the faction command systems in
\refsec{faction}, all four archetypes are defined but only the council has a
worked example. These limits motivate the early-stage framing: the contribution
is the engine and the contract.

\section{Conclusion}\label{sec:conclusion}
We presented \wopr{}, a simulation environment built on a replay-validated
rules engine, with a decision-point contract, a press ladder, and collective
faction command systems layered on that engine. We instantiated it here for one
wargame; the environment targets organizational decision-making more generally.

\bibliographystyle{colm2026_conference}
\bibliography{references_social_sim}

\section*{Ethics Statement}

\wopr{} simulates a commercial card game. \emph{Nuclear War} is a satirical product,
and our engine encodes its rules, not a model of real nuclear command,
deterrence, or crisis behavior. Agents make fictional in-game choices under
those rules. We make no fidelity claim for the real-world command systems
named as archetype origins in \reftab{c2origins}; they motivate the design and
are not modeled.

The work involves no human subjects and collects no human data. Agents are
language-model programs whose only outputs are legal game actions and, under
the press rungs, trace-only messages between in-game factions. Because speech
is recorded but not enforced, any analysis of betrayal, defection, or
escalation is a post-hoc read of recorded traces, not a behavioral claim about
deployed systems. We discourage extrapolating in-game outcomes to real
strategic decisions.

\section*{Reproducibility Statement}

The engine is deterministic: given a seed and an agent configuration, a game
produces one sequence of legal state transitions, and that sequence is the
artifact other researchers can re-derive. Every game is recorded as a replay
JSON that an independent validator checks against the rules, and the
parity invariant (\refapp{replay}) is enforced by a 240-game sweep run as part
of the standard gate.

All tooling runs through \texttt{uv} from the \texttt{nuclear\_war/}
directory. The full gate runs the pytest suite, requires clean \texttt{ruff}
and \texttt{pyright} output, requires \texttt{validate-rules} to return
\texttt{ok: true}, and runs the 240-game replay sweep. Checked-in example
configs cover each press rung, including a no-provider offline variant that
runs the full-press pipeline without a live model. A replay workbench
(\texttt{wopr\_visualizer}) loads replay JSON and the separate press-trace
sidecar for visual inspection. Full commands are in \refapp{repro}.

All code, example configurations, and replay artifacts are publicly available at
\url{https://github.com/eilab-gt/wopr}.

\section*{LLM Usage}

Following the COLM 2026 policy on the use of large language models, we
disclose LLM use in this work beyond minor assistance. Language models were
used in four roles.

\textbf{Implementation.} LLM coding agents wrote and revised substantial
portions of the engine, the agent harness, and the surrounding tooling in the
released repository. Every change was gated by the deterministic test suite,
the rules validator, and the replay parity sweep, and each was reviewed by a
human author before merging.

\textbf{Experiments.} LLM agents were used to drive experiment batches and the
model-calibration sweeps, and to assist in summarizing the recorded decision
and press traces.

\textbf{Writing.} LLM assistance was used in drafting and revising the prose of
this manuscript.

\textbf{Review.} LLM tools were used for adversarial review of the
implementation, for rules-fidelity audits against the published game rules, and
for automated review of pull requests.

This disclosure concerns tools used by the authors in producing the work. It is
separate from the language-model agents that are the object of study, which act
as players inside the environment and are described in the body of the paper.

The authors reviewed all of the above and take full responsibility for the
content, claims, and results of this paper.

\appendix
\section{Decision-Point Contract}\label{app:contract}

The engine exposes three entry points, in
\texttt{src/nuclear\_war\_env/decision\_loop.py} and \texttt{engine/decision.py}:

\begin{itemize}
  \item \texttt{pending\_decision(state) -> Decision | None}: the current
  choice, or \texttt{None} at a terminal state.
  \item \texttt{apply\_decision(state, action) -> list[EngineEvent]}: apply the
  chosen action, run mandatory steps until the next genuine choice, and store
  the resulting decision on the state.
  \item \texttt{observe(state, agent\_id) -> Observation}: a structured
  per-agent view. The table loop passes this into \texttt{choose}.
\end{itemize}

An agent is anything satisfying the protocol in
\texttt{agent\_protocol.py}:

\begin{verbatim}
class DecisionAgent(Protocol):
    def choose(self, observation: Observation,
               options: list[LegalAction]) -> LegalAction: ...
\end{verbatim}

A legacy adapter wraps older \texttt{choose(actions)} agents. Baseline agents
live in \texttt{baseline.py} (\texttt{HeuristicAgent}, \texttt{RandomAgent});
the observation-driven \texttt{ObservationHeuristicAgent} is exposed as the
\texttt{decision\_heuristic} seat.

A turn is a sequence of phases on a \texttt{DecisionCursor}: draw, secrets,
slide, propaganda, deterrents, place, attack, intercept, final strike, end.
Mandatory steps auto-advance; only genuine choices pause as a
\texttt{Decision}. The eight strategic decision types are: \texttt{PLACE},
\texttt{LAUNCH\_TARGET}, \texttt{SECRET\_TARGET},
\texttt{PROPAGANDA\_TARGET}, \texttt{FINAL\_STRIKE\_TARGET},
\texttt{INTERCEPT} (defender), \texttt{MODIFY\_DETERRENT}, and \texttt{PASS}.

The engine presents option lists in a fixed order: weakest opponent first for
launch targeting, highest population first for secret and propaganda targets,
and eligible anti-missiles before decline for interception. Observation
preserves hidden information: a defender is even offered an interception choice
when it has no anti-missile, so the option list never reveals the defender's
hand. The baseline heuristic picks \texttt{options[0]}, reproducing the
deterministic engine policy without consuming RNG draws. That behavior makes
the parity invariant in \refapp{replay} checkable.

\section{Replay Validation Method}\label{app:replay}

Given a seed, the deterministic engine can serialize every game to replay JSON
and re-execute it. We use replay validation as a fidelity method with two
anchors.

\paragraph{The parity invariant.}
Migrating a rule into the decision interface must preserve seeded table
outcomes byte-for-byte when the change is structural. When a fidelity fix
intentionally removes a model bug, affected seeded outcomes change; the test
suite asserts that divergence explicitly. For example, fixing duplicate
physical card ids changed a documented set of seeded outcomes; the pre-fix and
post-fix divergences are both pinned by name.

\paragraph{The 240-game sweep.}
The end-to-end gate runs the table simulation across 3 and 4 players, the
heuristic and random agents, and seeds 1--60:

\begin{verbatim}
for players in (3, 4):
    for agent in ("heuristic", "random"):
        for seed in range(1, 61):
            simulate --mode table --players players \
                     --seed seed --agent agent
\end{verbatim}

A migration that preserves parity keeps all 240 outcomes identical to the
prior head. The strongest check is a two-worktree diff. It runs baseline-agent
outcomes at the prior head and current head, normalizes the JSON round-trip
(tuples become lists), and diffs the results. The diff proves that structural
refactors leave outcomes unchanged, or it identifies exactly which seeded
outcomes changed for a documented fidelity fix. It does not depend on a single
hardcoded golden.

Random-agent outcomes diverge by design wherever targeting or interception is a
real decision, because the random agent draws RNG to choose there. That
divergence is asserted by name in
\texttt{test\_random\_table\_outcomes\_diverged\_from\_a0} rather than absorbed
silently.

\section{Demo Visualizations}\label{app:demo-visualizations}

These screenshots show the local replay workbench, batch analysis view, and
live workbench in the paper theme. They are UI captures used for visual
inspection and discussion, not additional experimental measurements.

\begin{figure*}[p]
\centering
\includegraphics[width=\textwidth]{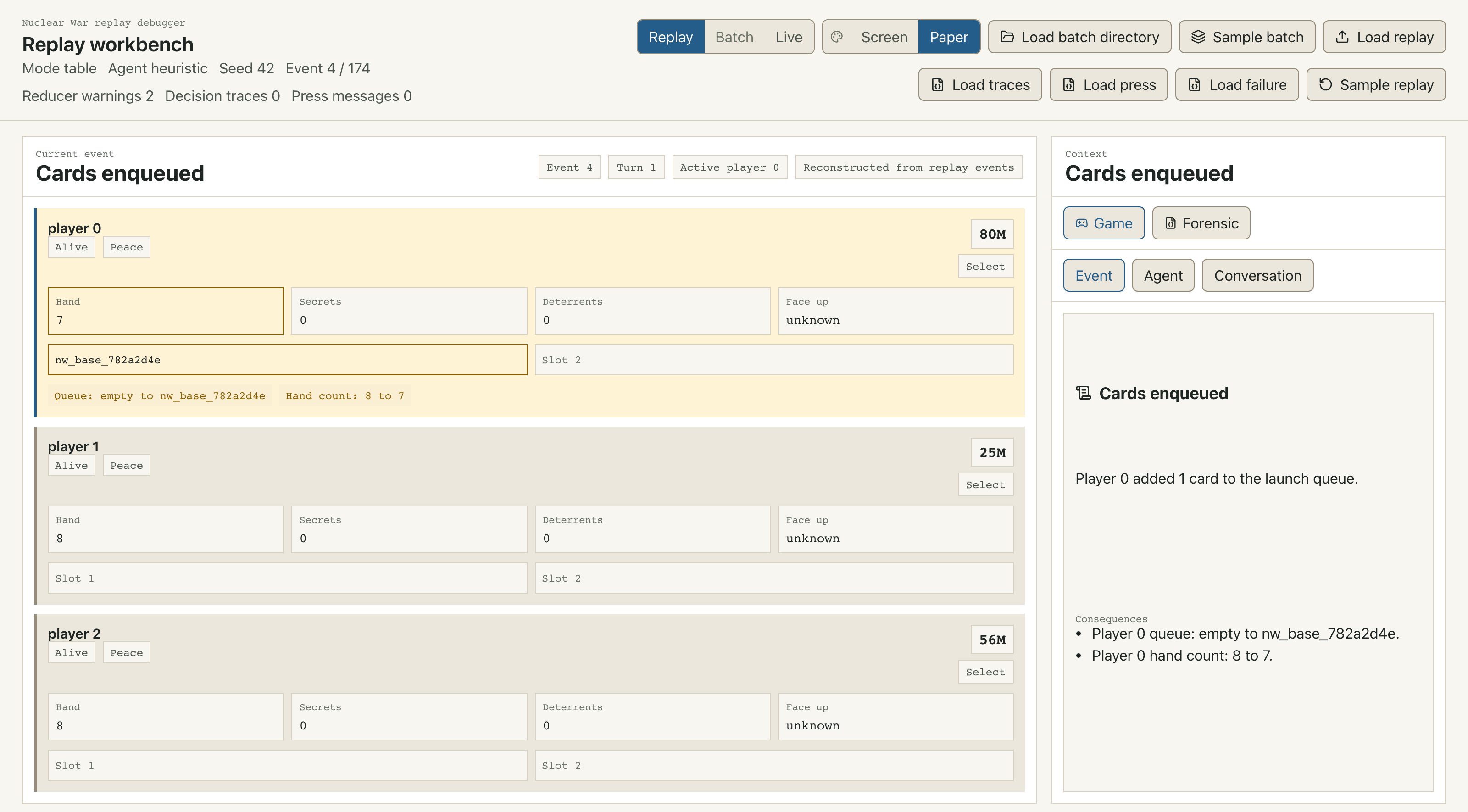}
\caption{Replay workbench table view for a cards-enqueued event.}
\label{fig:demo-replay-table-event}
\end{figure*}

\begin{figure*}[p]
\centering
\includegraphics[width=\textwidth]{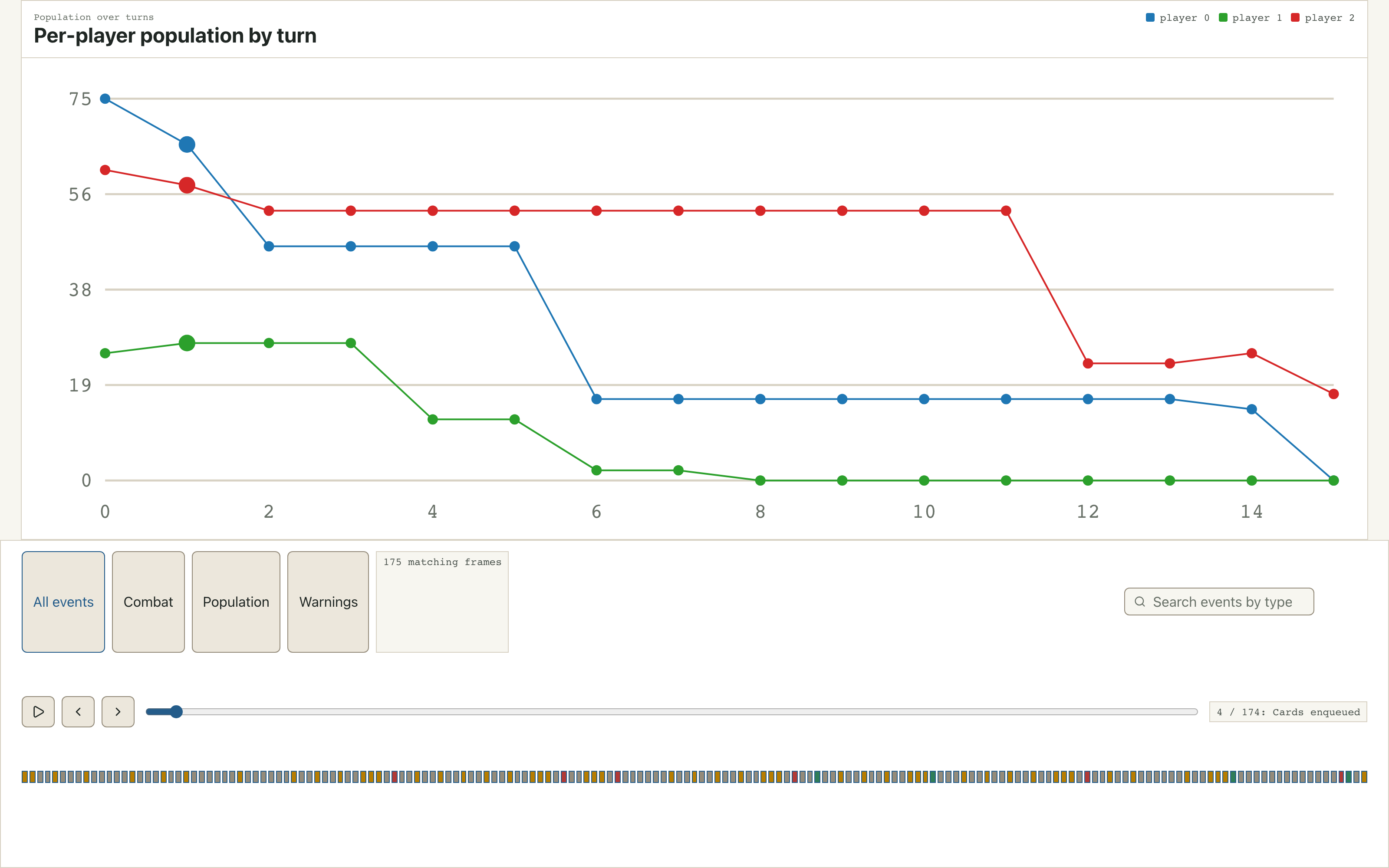}
\caption{Population chart and event timeline for the same replay. At turn~0
the top line is player~0, the middle player~2, and the bottom player~1.}
\label{fig:demo-population-timeline}
\end{figure*}

\begin{figure*}[p]
\centering
\includegraphics[width=0.55\textwidth]{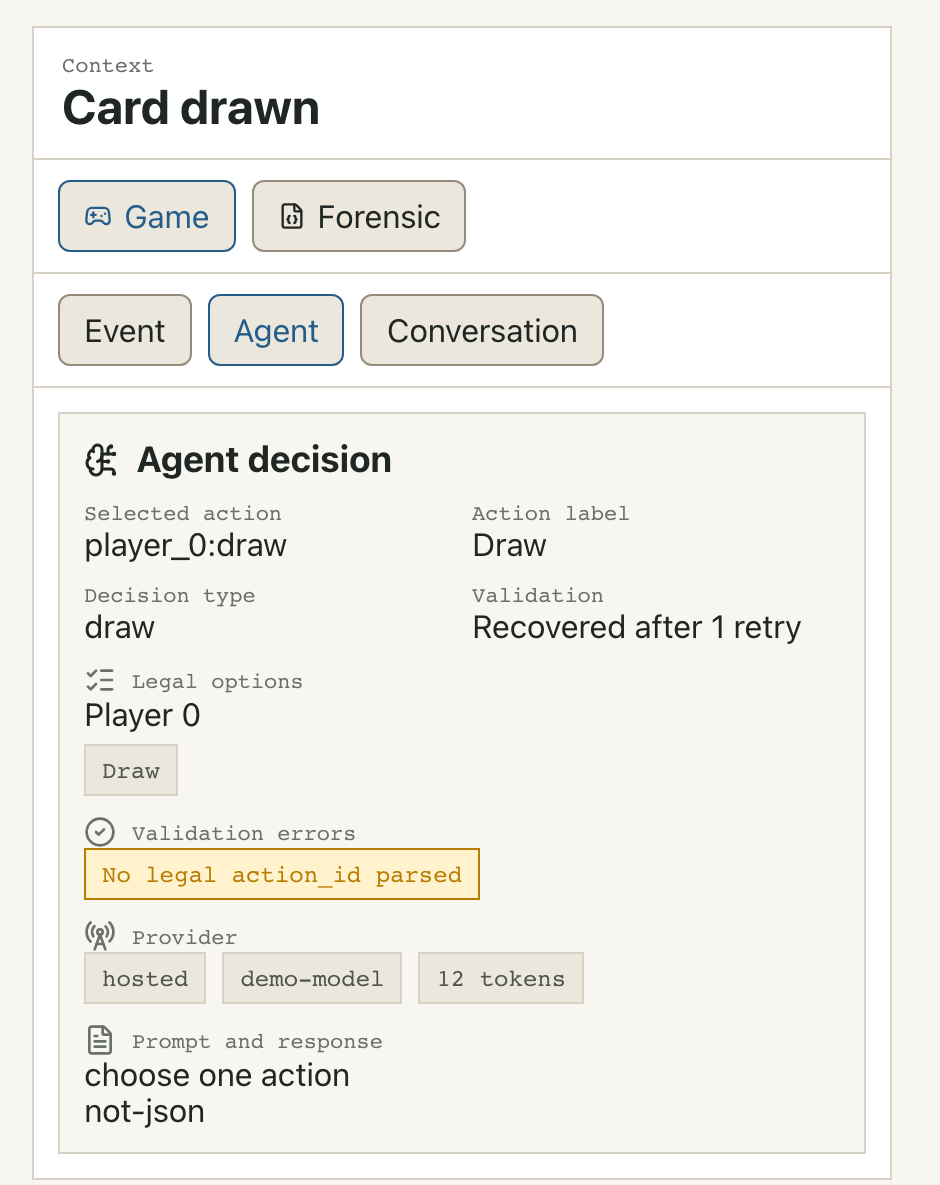}
\caption{Decision-trace inspector showing the selected action context. The
trace shown records a malformed (non-JSON) model response recovered after one
retry.}
\label{fig:demo-agent-decision-trace}
\end{figure*}

\begin{figure*}[p]
\centering
\includegraphics[width=0.55\textwidth]{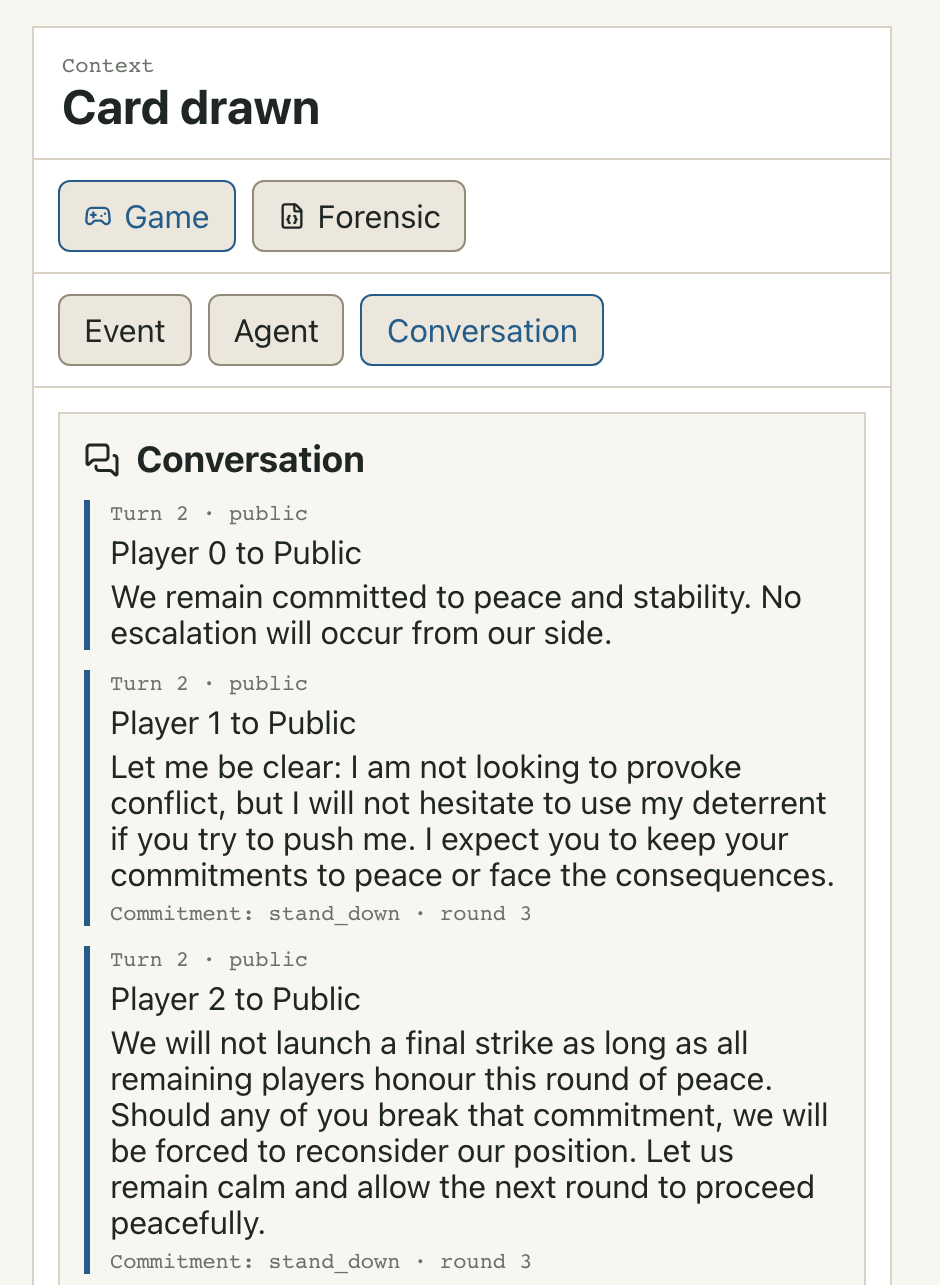}
\caption{Conversation inspector for a full-press game. The public statements
shown carry structured \texttt{stand\_down} commitments for round~3.}
\label{fig:demo-conversation-press}
\end{figure*}

\begin{figure*}[p]
\centering
\includegraphics[width=0.55\textwidth]{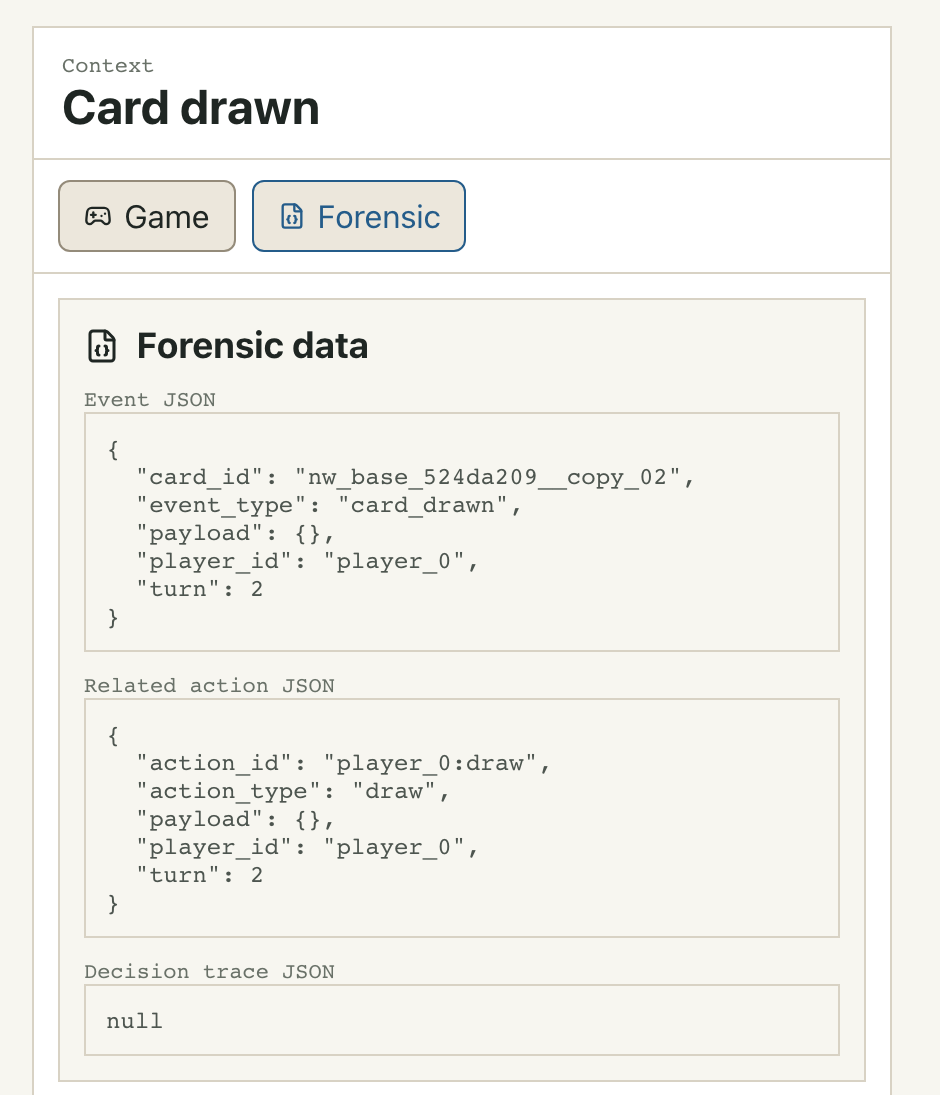}
\caption{Forensic inspector exposing replay JSON for audit.}
\label{fig:demo-forensic-json}
\end{figure*}

\begin{figure*}[p]
\centering
\includegraphics[width=\textwidth]{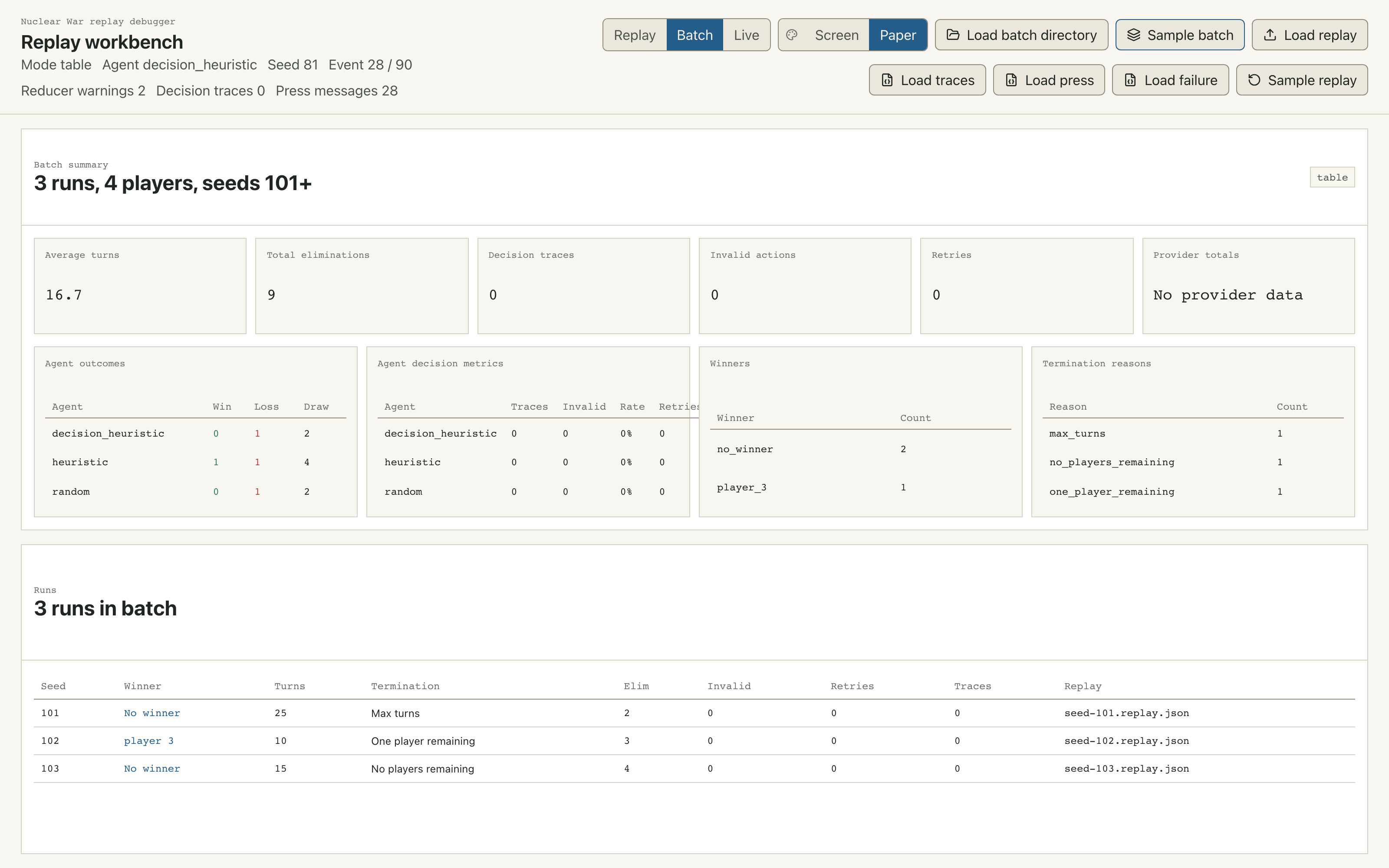}
\caption{Batch analysis view for the bundled three-run sample.}
\label{fig:demo-batch-analysis}
\end{figure*}

\begin{figure*}[p]
\centering
\includegraphics[width=\textwidth]{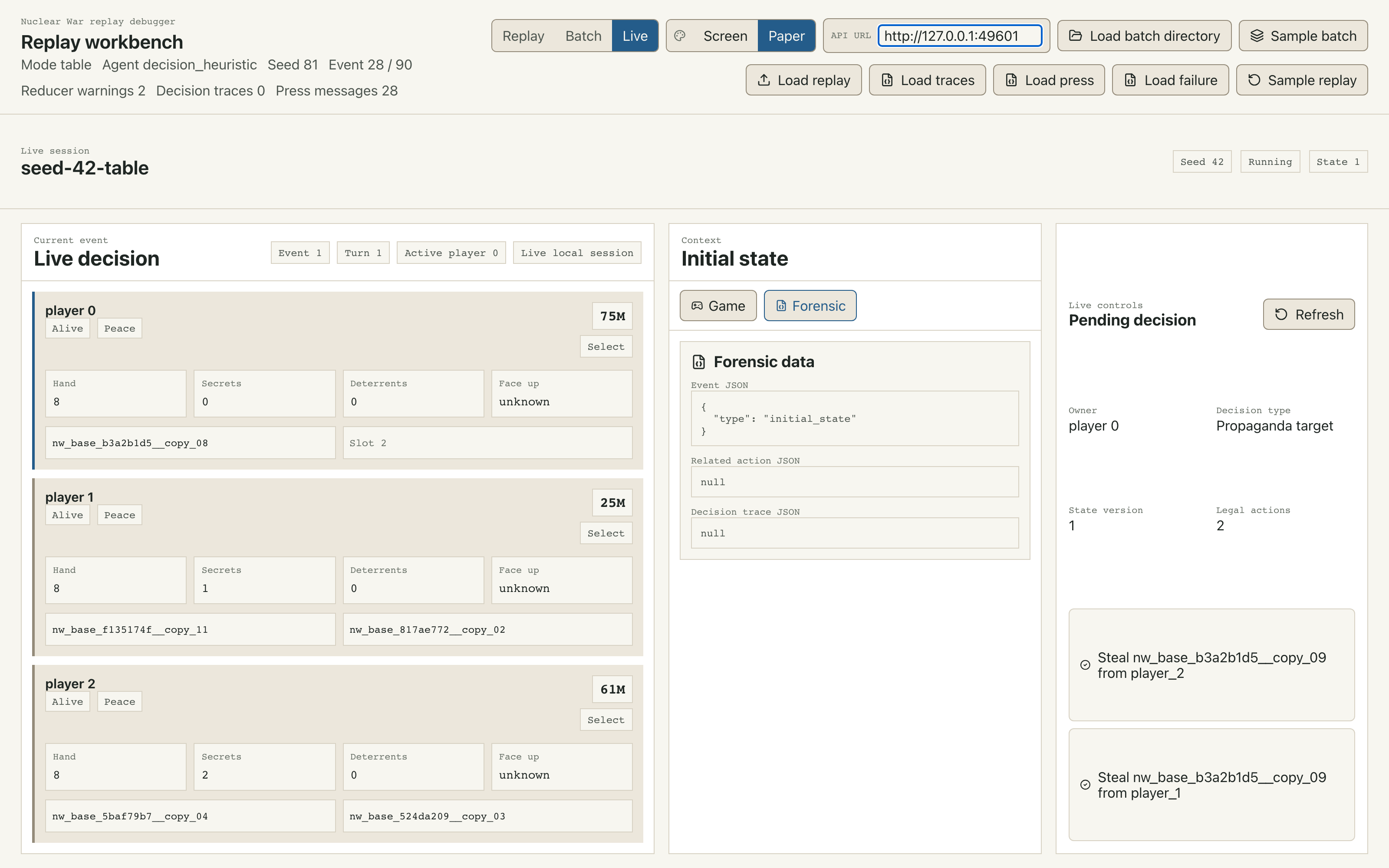}
\caption{Live workbench connected to the local rules API.}
\label{fig:demo-live-play}
\end{figure*}

\clearpage

\section{Rule Conformance}\label{app:conformance}

We do not claim a one-to-one reproduction of the published rulebook. Instead
the engine keeps a rule conformance matrix that traces each implemented
mechanic to a source, and a rules-trace mechanism that maps a full
deterministic table replay onto source-linked rule-step ids and reports gaps.

\paragraph{Traced mechanics.}
The matrix records the active edition variant, the base-card registry, and
the implemented effects for warheads, delivery systems, propaganda,
anti-missiles, secrets, the fallout spinner, final strike, and the postal
equipment families. Each implemented effect is bound to a registry record and
a source note. The rules-trace summary maps every replay action and event in
one deterministic full table game to a source-mapped rule step and reports no
full-game trace gaps.

\paragraph{Tracked limits.}
The matrix also records simplifications we have not resolved. Two matter for
reproducibility. First, postal equipment launch resolution (space platforms,
cruise missiles, killer satellites, Space Shuttle attacks, and the MX
missile's per-segment rolls) uses the base two-d10 fallout spinner where the
postal rules specify a separate six-sided Radioactive Fallout die; the
success and failure families are preserved, but the launch-failure
probability differs. Second, exact card text, expansion deck composition,
alternate editions, and postal press adjudication are not verified. These are
tracked simplifications, not source-backed fidelity choices.

\section{Press-Ladder Mechanics}\label{app:pressmech}

\begin{figure}[t]
\centering
\includegraphics[width=0.85\linewidth]{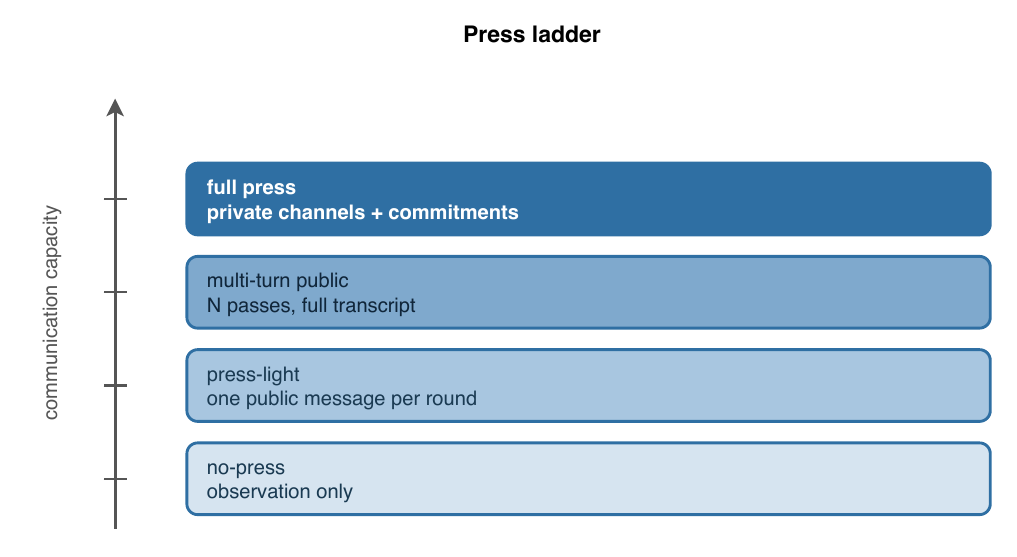}
\caption{The press ladder as a communication-capacity axis on an unchanged
rules engine. Rungs increase what players may say to one another, from silence
to private single-recipient channels with structured commitments. Speech never
mutates game state.}
\label{fig:press-ladder}
\end{figure}

Across all rungs, press is trace-only: it never mutates \wopr{} game state, and
the replay JSON is byte-identical to a no-press game. The four rungs differ in
what an agent may say and to whom (\reffig{press-ladder}).

\paragraph{Per-rung option sets.}
\emph{No-press} offers no communication options. \emph{Press-light} gives each
living agent one public message per round, which the system injects into
agents' memory for the next round. \emph{Multi-turn public} runs $N$ fixed
passes per round. Each living agent speaks once per pass and sees the
accumulating transcript. \emph{Full press} adds three choices per pass:
\texttt{decline}, \texttt{speak} (public), and \texttt{whisper} (private, to
one named recipient). When an agent chooses \texttt{whisper}, it specifies a
\texttt{to} recipient; the parser extracts the recipient and an optional
commitment from the model's JSON response.

\paragraph{The visibility rule.}
A privacy filter \texttt{visible\_to(player, transcript)} determines what each
agent sees. A public message is visible to all living players. A private
whisper is visible only to its sender and its single recipient. For each
message, the recorded \texttt{prior\_messages} stores the filtered view that
the speaker saw; a separate omniscient sidecar records every message in full
for the researcher.

\paragraph{Commitment schema.}
A speaker may attach a structured commitment to a public statement or a
private whisper. The schema is:
\begin{verbatim}
{"kind": "stand_down",
 "target_round": 3,            # optional
 "notes": "details"}           # optional
\end{verbatim}
The system records commitments as data and links them to later decisions
through \texttt{linked\_decision\_traces} when applicable. The system performs
no automatic violation detection; researchers analyze violations against the
post-game artifact.

\paragraph{Parse robustness.}
A recoverable-once retry policy handles malformed model output (invalid JSON or
an invalid recipient). In strict mode, repeated parse failures fail the run
hard rather than silently degrading the transcript.

\section{Faction Command-and-Control: Detail}\label{app:factionc2}

\begin{figure*}[t]
\centering
\includegraphics[width=0.95\linewidth]{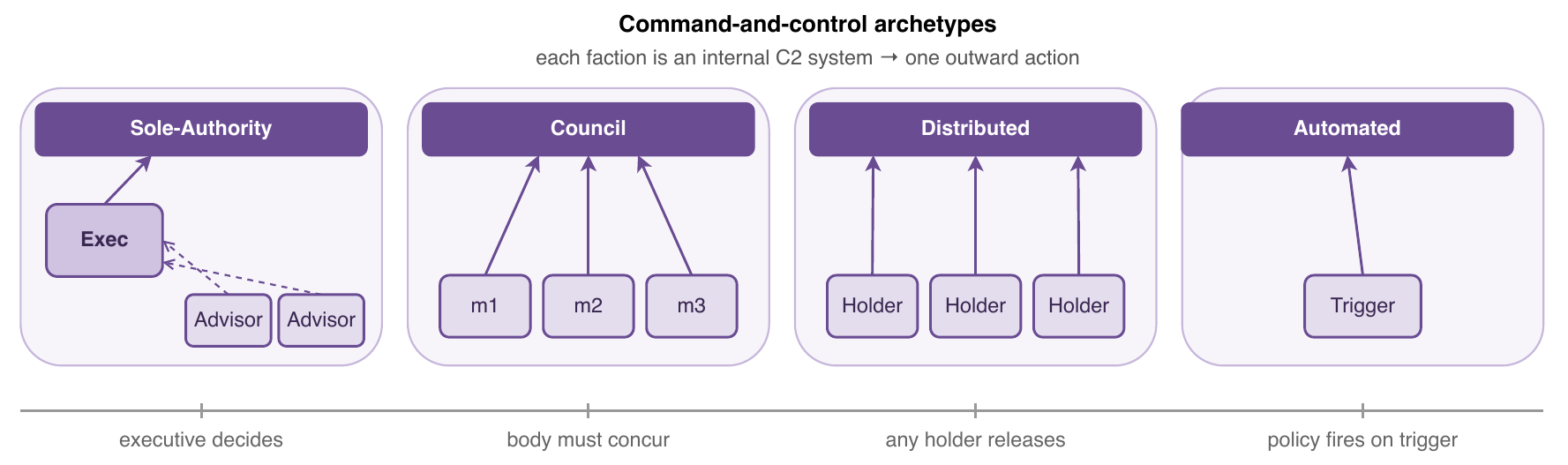}
\caption{Four command-and-control archetypes a faction may instantiate,
distinguished by where release authority aggregates. Sole-authority places
authority in one executive who may ignore advisors. A council must concur and
binds its members. Distributed release lets any authorized holder act.
Automated release fires a pre-armed policy on a trigger without fresh
deliberation. All four return one action per decision, so they are directly
comparable as seats in the same game despite their structural differences.
Real-world command systems cited as archetype origins in \reftab{c2origins}.}
\label{fig:faction-c2}
\end{figure*}

A \texttt{FactionDecisionAgent} builds a composite faction with one subordinate
client per member. At each decision, the faction collects one vote per member,
applies the archetype's aggregation rule, and returns one \texttt{LegalAction}.
The engine sees a single action. The agent records the deliberation in
\texttt{last\_deliberation}: the vote list, the rule applied, and the selected
action.

\paragraph{Aggregation rules.}
Each archetype maps its members' votes to one action via a pure
aggregation function. We demonstrate the binding case: councils.

\begin{description}
\item[Sole-authority] returns the executive's action, or defers to the
first advisor when a \texttt{deference} parameter is saturated. Advisors
cannot bind the executive.
\item[Council] accumulates per-member weights per action id. The
highest-scoring action wins if its share of total weight meets a
\texttt{threshold}; otherwise the faction defaults to the first-cast vote.
Ties break deterministically toward the earliest-cast action, so a seeded game
replays exactly.
\item[Distributed] treats any non-default (non-pass) vote as a release
authorization. If at least a \texttt{quorum} casts distinct release actions,
the first release action wins; otherwise the faction holds.
\item[Automated] returns the \texttt{policy\_action\_id} set when the policy
was armed; it reads no member votes at release time.
\end{description}

\paragraph{Configurable parameters.}
The structure fixes what is configurable; \reftab{c2params} lists the
parameter each archetype exposes.

\begin{table}[h]
\centering
\small
\begin{tabular}{lll}
\toprule
Archetype & Parameter & Effect \\
\midrule
sole-authority & \texttt{deference} & 0: ignore staff; 1: defer to first advisor \\
council        & \texttt{weights}   & per-member vote weight \\
council        & \texttt{threshold} & share of weight to win (default 0.5) \\
distributed    & \texttt{quorum}    & distinct releases required to act \\
automated      & \texttt{policy\_action\_id} & required pre-armed action (no default) \\
\bottomrule
\end{tabular}
\caption{Per-archetype configuration. All are seat-config fields named
under the \texttt{faction\_c2} agent in the experiment config.}
\label{tab:c2params}
\end{table}

\paragraph{Worked trace: a council release decision.}
\reftab{c2trace} shows one \texttt{LAUNCH\_TARGET} decision by a
three-member council (archetype \texttt{council}, threshold $0.34$). The
faction observes a two-opponent endgame; members disagree. Member rationales
are illustrative prose matched to the recorded votes. The vote ids,
aggregation rule, and returned action come from a real
\texttt{last\_deliberation} record. Because no option clears the threshold,
the rule returns \texttt{threshold\_not\_met\_default}. The faction then
defaults to the first-cast vote, which becomes the single action recorded by
the engine.

\begin{table}[h]
\centering
\small
\begin{tabular}{lll}
\toprule
Member & Vote & Stated rationale (illustrative) \\
\midrule
land-command & \texttt{target p2} & p2 is depleted; finish them \\
strategic-advisor & \texttt{pass} & conserve; p3 is the larger threat \\
air-command & \texttt{target p3} & p3 holds more population \\
\midrule
\multicolumn{3}{l}{\textbf{Rule:} \texttt{threshold\_not\_met\_default}} \\
\multicolumn{3}{l}{\textbf{Faction action:} \texttt{target p2} (first-cast vote)} \\
\bottomrule
\end{tabular}
\caption{One council deliberation over a \texttt{LAUNCH\_TARGET} decision.
Votes and the aggregation outcome are from a real \texttt{last\_deliberation};
rationale prose is illustrative. The faction returns one action to the engine,
linked to the replay like any agent action.}
\label{tab:c2trace}
\end{table}

\paragraph{Literature origins.}
\reftab{c2origins} maps each archetype to a real nuclear-use command system
that motivates it. These entries provide origin and motivation, not fidelity
targets; the simulation does not claim to model any named state.

\begin{table}[h]
\centering
\small
\begin{tabular}{lp{0.62\linewidth}}
\toprule
Archetype & Real-world origin (cited) \\
\midrule
sole-authority & US formal presidential release authority$^{1}$
\citep{crs_if10521_launch} \\
council        & Pakistan NCA, a civil-military council authority
\citep{pakistan_nca_act2010}; supporting accounts
\citep{khan2012eatinggrass,salik2017learning,kidwai2002thresholds},
comparative survey \citep{lewis2019fingerbutton}$^{2}$ \\
distributed    & Contingent pre-delegation of use authority
\citep{feaver2017predelegation} \\
automated      & Semi-automated retaliatory assurance (Soviet Perimeter)
\citep{hoffman2009deadhand,yarynich2003c3}$^{3}$ \\
\bottomrule
\end{tabular}
\caption{Each archetype is grounded in a real nuclear-use command
structure, cited as its origin. The four archetypes are analytic
abstractions inspired by recurring C2 design dimensions in the literature;
they do not map one-to-one to the automaticity, devolution, delegation,
and pre-delegation mechanisms of \citet{feaver2017predelegation}.
Permissive action links are a separate negative-control, use-denial
mechanism and are not modeled here as a release-authority archetype.
\textit{Notes:} $^{1}$Under normal procedure; ``sole authority'' does not
exclude historical advance-authorization \citep{frus1957_nsc5602}.
$^{2}$Public primary sources establish the NCA's statutory structure, not
a real-time launch voting rule. $^{3}$Human activation and a human launch
decision remained part of the known accounts; not a fully autonomous
launch authority.}
\label{tab:c2origins}
\end{table}

\paragraph{Parity.}
A faction is a \texttt{DecisionAgent} that returns one \texttt{LegalAction}
per \texttt{choose} call. The engine's decision and apply path is
unchanged: the table runner calls \texttt{choose}, validates the single
returned action against the legal options, and applies it. Because the
faction is a pure aggregation layer over deterministic member agents,
same-seed determinism is preserved and is covered by
\texttt{test\_faction\_replay\_determinism} (two runs of one seeded
faction game produce an identical replay). Deliberation lives in
\texttt{last\_deliberation} on the agent instance, not in the trace
sidecar, so the trace schema (v4) is untouched.

\section{Concordia Connector Architecture}\label{app:connector}

The Concordia integration has a one-way dependency: the connector package
(\texttt{nuclear\_war\_concordia}) imports from Concordia, but \wopr{} never
does. The two systems touch through one optional \texttt{press\_hook}. This
design keeps \wopr{} a standalone rules engine that any harness can drive, and
keeps the press ladder as a layer the engine can ignore.

The harness seats Concordia-style agents (identity, role, objective) at the
\wopr{} table and routes their decisions through the decision-point contract
(\refapp{contract}). A generic OpenAI-compatible HTTP client produces press, rather than
native Concordia press machinery, so the press producer remains replaceable.
Because the press layer writes only to a sidecar and never mutates \wopr{}
replay JSON, the active press rung does not affect engine parity or replay
validation.

\section{Reproducibility}\label{app:repro}

Commands below assume \texttt{uv} from the \texttt{nuclear\_war/} directory.

\paragraph{Running a full-press game.}
The full-press configuration is the checked-in example
\texttt{docs/examples/concordia\_full\_press\_together\_demo.json} (four
\texttt{concordia\_http} seats, \texttt{full\_press} mode, two passes, five
turns). The offline no-provider variant
\texttt{concordia\_full\_press\_style\_demo.json} uses
\texttt{concordia\_first\_legal} seats. It produces decline-only traces and
proves the pipeline runs without a live model.

\paragraph{Other rungs.}
Press-light and multi-turn public have analogous example configs. The press
modes are single-game (\texttt{concordia-demo}); there is no batch runner for
press.

\paragraph{Visualizer.}
The \texttt{wopr\_visualizer} replay workbench loads replay JSON plus the
separate press-trace sidecar and renders the Conversation tab. Private
whispers render distinctly (``speaker whispered to recipient'') and commitment
badges appear when present.

\end{document}